\documentclass[aps,prl,reprint,longbibliography,notitlepage,superscriptaddress]{revtex4-1}
\usepackage{graphicx,tikz}
\usepackage{amssymb,mathrsfs}
\usepackage{bm}
\usepackage{slashed}
\usepackage{dcolumn}
\usepackage{hyperref}
\usepackage{amsmath}
\usepackage{float}
\usepackage{dsfont}
\usepackage[utf8]{inputenc}

\begin{document}

\title{Extracting many-body correlators of saturated gluons with precision from inclusive photon+dijet final states in deeply inelastic scattering}

\author{Kaushik Roy}
\email{kaushik.roy.1@stonybrook.edu}
\affiliation{Department of Physics and Astronomy, Stony Brook University, Stony Brook, NY 11794, USA}
\affiliation{Physics Department, Brookhaven National Laboratory, Bldg. 510A, Upton, NY 11973, USA}

\author{Raju Venugopalan}
\email{raju@bnl.gov}
\affiliation{Physics Department, Brookhaven National Laboratory, Bldg. 510A, Upton, NY 11973, USA}

\date{\today}

\begin{abstract}
We highlight the principal results of a computation~\cite{Roy:2019hwr} in the Color Glass Condensate effective field theory (CGC EFT) of the next-to-leading order (NLO) impact factor for inclusive photon+dijet production at Bjorken $x_{\rm Bj} \ll 1$ in deeply inelastic electron-nucleus (e+A DIS) collisions. When combined with extant results for next-to-leading log $x_{\rm Bj}$ JIMWLK renormalization group (RG) evolution of gauge invariant two-point (``dipole") and four-point (``quadrupole") correlators of light-like Wilson lines, the inclusive photon+dijet e+A DIS cross-section can be determined to $\sim 10$\% accuracy. Our computation simultaneously provides the ingredients to compute fully inclusive DIS, inclusive photon, inclusive dijet and inclusive photon+jet channels to the same accuracy. This makes feasible quantitative extraction of many-body correlators of saturated gluons and precise determination of the saturation scale $Q_{S,A}(x_{\rm Bj})$ at a future Electron-Ion Collider. An interesting feature of our NLO result is the structure of the violation of the soft gluon theorem in the Regge limit. Another is the appearance in gluon emission of time-like non-global logs which also satisfy JIMWLK RG evolution.  
\end{abstract}

\maketitle

The many-body recombination and screening of gluons in the high energy or small Bjorken $x_{\rm Bj}$ Regge limit of QCD competes with their rapid bremsstrahlung and leads to the perturbative unitarization of cross--sections. This gluon saturation phenomenon~\cite{Gribov:1984tu,Mueller:1985wy}, in the Color Glass Condensate (CGC) effective field theory (EFT) picture~\cite{McLerran:1993ni,McLerran:1993ka,McLerran:1994vd,Iancu:2003xm,Gelis:2010nm,Kovchegov:2012mbw,Blaizot:2016qgz}, occurs when the phase-space occupancy $n$ of gluons for transverse momenta $k_\perp \leq Q_S(x_{\rm Bj})$ becomes of the order of the inverse of the QCD coupling $\alpha_S$. The saturation scale $Q_S(x_{\rm Bj})$ is an emergent quantity and is the only large scale in the Regge limit; because it is much larger than intrinsic QCD scales, asymptotic freedom dictates that $\alpha_S(Q_S)\ll 1$.  The large mode occupancy $n \sim 1/\alpha_S(Q_S) \gg 1$ therefore suggests that  gluon saturation corresponds to a remarkable classicalization of QCD at high energies.  


In this letter, we will discuss the first computation in the CGC EFT of the next-to-leading order (NLO) "impact factor" for inclusive photon+dijet production in deeply inelastic scattering of electrons off nuclei (e+A DIS) at high energies. A powerful motivation to perform the computation is the prospect of such measurements~\footnote{Our highly differential computation will allow us to provide, to the same accuracy, cross-sections for inclusive dijet, inclusive photon, inclusive photon and fully inclusive DIS measurements.} at a future Electron-Ion Collider (EIC)~\cite{Accardi:2012qut,Aschenauer:2017jsk}.
As we will outline here, knowing the NLO impact factor will enable us to compute the photon+dijet cross-section in e+A DIS to O($\alpha_S^3 \ln(1/x_{\rm Bj})$) accuracy. The details of the computation are spelled out in a companion paper~\cite{Roy:2019hwr}.  At the energies and nuclear saturation scales accessible at an EIC, the computations allow predictions to $\sim 10$\% accuracy~\footnote{An important caveat is that this statement is modulo the scale and scheme dependence of the results, which remain to be quantified.}. This may be sufficient for precision tests of the CGC EFT and to distinguish its predictions from those of potential alternative frameworks. 

It is instructive to first briefly consider our previous computation~\cite{Roy:2018jxq} in the CGC EFT of the leading order (LO) inclusive photon+dijet ($\gamma+q{\bar q}$) e+A DIS cross-section:
\begin{equation}
\frac{\mathrm{d}^{3} \sigma^{{\rm LO};\gamma+q\bar{q}+X}}{\mathrm{d}x \,  \mathrm{d}Q^{2} \mathrm{d}^{6} K_{\perp} \mathrm{d}^{3} \eta_{K} }= \frac{\alpha_{em}^{2}q_{f}^{4}y^{2}N_{c}}{512 \pi^{5} Q^{2}} \, \frac{1}{(2\pi)^{4}} \,  \frac{1}{2} \,  L^{\mu \nu} \tilde{X}_{\mu \nu}^{\text{LO}} \, .
\label{eq:triple-differential-CS-LO}
\end{equation}
Here $\alpha_{em}=e^{2}/4\pi$ is the electromagnetic fine structure constant,  $y=q\cdot P_{N}/\tilde{l}\cdot P_{N}$ is the inelasticity, $Q^{2}= -q^{2} >0$ is the squared momentum transfer from the nucleus, and $\mathrm{d}^{6} K_{\perp} \mathrm{d}^{3} \eta_{K}$ collectively denotes the phase space density of final state quark, anti-quark and photon. Likewise, $L^{\mu \nu}$ is the lepton tensor, corresponding to the emission of a virtual photon with four momentum $q^\mu$ by the electron~\footnote{Explicit expressions for all kinematic variables are provided in \cite{Roy:2018jxq} and in \cite{Roy:2019hwr}.}. Our focus here is on the scattering of the virtual photon on the nuclear target producing the $\gamma+q\bar{q}$ final state and other (phase space integrated) $X$ particles, described by the LO hadron tensor, 
\begin{equation}
\tilde{X}_{\mu \nu}^{\text{LO}}= \int [\mathcal{D} \rho_{A} ] \, W_{\Lambda_{0}^{-}} [\rho_{A}]\, \hat{X}_{\mu \nu}^{\text{LO}} [\rho_{A}]
\, .
\label{eq:leading-order-dsigma}
\end{equation}
In a Born-Oppenheimer separation of modes in the EFT, the $\rho_{A}$ are the large $x_{\rm Bj}$ static color sources in the nucleus; these correspond to light cone longitudinal momentum modes with $\Lambda^- < \Lambda_{0}^{-}$. The initial distribution of such modes at the scale $\Lambda_{0}^{-}$ is given by the nonperturbative gauge invariant weight functional $W_{\Lambda_{0}^{-}} [\rho_{A}]$. The small $x_{\rm Bj}$ dynamical gluon fields interacting with the probe correspondingly have longitudinal momenta above $\Lambda_{0}^{-}$; the leading order classical gluon field is a ``shock wave" solution of the Yang-Mills (YM) equations in the presence of the sources $\rho^a (\bm{x}) = {\tilde \rho}^a(\bm{x}_\perp)\delta(x^-)$~\footnote{The delta function corresponds to the Lorentz contraction of the sources in a frame where the momentum of the right moving nucleus, $P^{+}_{N}  \rightarrow \infty$.}  of O($1/g$). 

The solution of the YM equations in the Lorenz gauge $\partial_\mu A^\mu=0$ (or equivalently light cone (LC) $A^-=0$ gauge) is given by 
\begin{align}
& A_{\rm cl}^+ = \int \frac{\mathrm{d}^2 \bm{z}_\perp}{ 4\pi} \ln\frac{1}{(\bm{x}_\perp-\bm{z}_\perp)^2\Lambda^2}\, \rho_A(x^-,\bm{z}_\perp) \,,\nonumber\\
& A_{\rm cl}^{-}=0\,\,;\, \, A_{{\rm cl},\perp} =0\,,
\label{eq:A+}
\end{align}
where $\Lambda$ is an infrared cutoff  necessary to invert the Laplace equation $-\nabla_\perp^2 A_{\rm cl}^+ = g\rho_A$. This solution in Lorenz gauge is simply related to the solution in the LC gauge ${\tilde A}^+=0$, with ${\tilde A}_{\rm cl}^-=0$ and ${\tilde A}_{\rm cl}^i = \frac{i}{g} U\partial^i U^\dagger$, where the adjoint lightlike Wilson line 
\begin{equation}
U(\bm{x}_{\perp})=P_{-} \Bigg( \text{exp} \Bigg\{ -ig \int_{-\infty}^{+\infty} \mathrm{d}z^{-} A_{\rm cl}^{+,a} (z^{-},\bm{x}_{\perp}) \,T^a\Bigg\} \Bigg) \, .  
\label{eq:Wilson-line}
\end{equation}
is expressed in terms of the the large $x$ static color source densities via Eq.~\ref{eq:A+}. Note that $T^a$, $a=1,\cdots,8$,  are the generators of color $SU(3)$ in the adjoint representation. This $x^-$ path ordered exponential, and its fundamental representation counterpart ${\tilde U}(\bm{x}_{\perp})$ (obtained by replacing $T^a$ with the Gell-Mann matrices $t^a$), efficiently resum all higher twist contributions $\frac{\rho_{A}}{\nabla_\perp^2}\rightarrow \frac{Q_S}{Q^2}$ from the multiple scattering of the $q\bar{q}$ pair off the color field of the nucleus. 

\begin{figure}[!htbp]
\begin{minipage}[b]{0.3\textwidth}
\includegraphics[width=\textwidth]{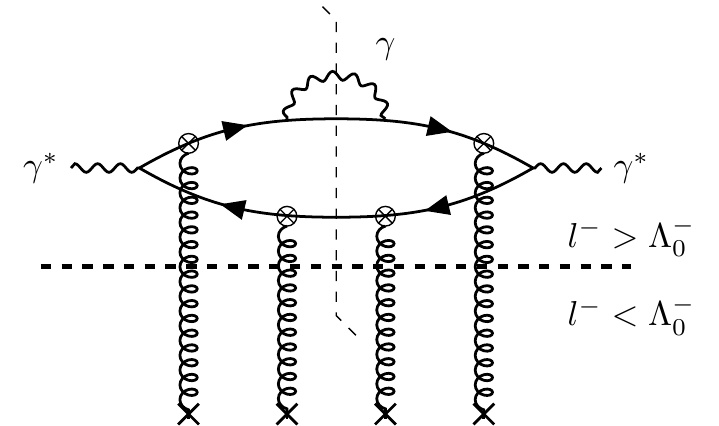}
\end{minipage}
\caption{A representative LO diagram. The cross-hatched open blobs represent the dressed quark propagator in $A^-=0$ gauge. See text for details. \label{fig:one}}
\end{figure}

The LO $\gamma+q\bar{q}$ amplitude is obtained by solving the Dirac equation in the $A_{\rm cl}^+$ shock wave background in the $A^-=0$ LC gauge. A Feynman diagram for the LO process is shown in 
Fig.~\ref{fig:one}, where the vertical dashed line represents the cut separating the amplitude from its complex conjugate amplitude and the horizontal dashed line separates the dynamical projectile modes from the static target shock wave gauge fields at the scale $\Lambda_0^-$. The dressed shock wave propagator, denoted by cross-hatch circles in the figure, has the remarkably simple solution~\cite{McLerran:1994vd,McLerran:1998nk},
\begin{equation}
S_{ij}(p,q) =  S_0(p)\,\mathcal{T}_{q;ij}(p,q)\,S_0 (q)\,,
\label{eq:dressed-quark-mom-prop}
\end{equation}
where $S_0 (p) = \frac{i \slashed{p}}{p^{2}+i \varepsilon}$ is the free massless fermion propagator, and 
\begin{align}
 \mathcal{T}_{q;ji}(q,p) &= 2 \pi \, \delta(p^{-}-q^{-}) \gamma^{-} {\text{sign} (q^-)} \nonumber \\
& \times \int \mathrm{d}^{2} \bm{z}_{\perp} \enskip e^{-i(\bm{q}_{\perp} - \bm{p}_{\perp})\cdot \bm{z}_{\perp}} \tilde{U}_{ji}^{\text{sign}(q^-)}(\bm{z}_{\perp}) \enskip ,
\label{eq:effective-quark-vertex}
\end{align}
is the effective vertex corresponding to the multiple scattering of the quark or antiquark off the shock wave~\footnote{The $\mathds{1}$ term in the expansion of the Wilson lines corresponds to the possibility that either the quark or the antiquark does not scatter. Since at least one of them must, one should subtract a term from the net amplitude wherein ${\tilde U}$ (and $U$) are set to unity everywhere.}. The LO computation of the $\gamma+q\bar{q}$ cross-section in $A^-=0$ gauge 
is straightforward and one finds,
\begin{align}
\tilde{X}^{\text{LO}}_{\mu \nu}&=  2\pi  \, \delta(1-z_{q}-z_{\bar{q}}-z_{\gamma}) \!\! \int \mathrm{d} \Pi_{\perp}^{\text{LO}} \!\! \int  \! \mathrm{d} {{\Pi_{\perp}^{\prime \text{LO}}}}^{\star} \!  \tau^{q\bar{q},q\bar{q}}_{\mu \nu}(\bm{l}_{\perp},\bm{l'}_{\perp}) \nonumber \\
& \times  \Xi(\bm{x}_{\perp},\bm{y}_{\perp};\bm{y'}_{\perp},\bm{x'}_{\perp}) \, ,
\label{eq:H-tensor-LO}
\end{align}
where we introduced the compact notation~\footnote{The second such term appearing in Eq.~\ref{eq:H-tensor-LO} results from the complex conjugate of Eq.~\ref{eq:transverse-phases-LO} and corresponds to replacing all transverse coordinates and internal momenta therein by their primed counterparts.}, 
\begin{equation}
\int \mathrm{d} \Pi^{\text{LO}}_{\perp} = \int_{\bm{l}_{\perp}} \int_{\bm{x}_{\perp},\bm{y}_{\perp}} e^{i \bm{l}_{\perp}.(\bm{x}_{\perp}-\bm{y}_{\perp})-i(\bm{k}_{\perp}+\bm{k}_{\gamma \perp}).\bm{x}_{\perp}-i\bm{p}_{\perp}.\bm{y}_{\perp}} \, .
\label{eq:transverse-phases-LO}
\end{equation}
The function 
$\tau^{q\bar{q},q\bar{q}}_{\mu \nu}(\bm{l}_{\perp},\bm{l'}_{\perp})$
denotes the spinor trace in the LO cross-section~\cite{Roy:2018jxq}. 

The nonperturbative input from the dynamics of saturated gluons in the nuclear target is contained in 
\begin{equation}
\Xi(\bm{x}_{\perp},\bm{y}_{\perp};\bm{y'}_{\perp},\bm{x'}_{\perp})=1-D_{xy}-D_{y'x'}+Q_{y'x';xy} \, .
\label{eq:LO-cross-section-color-structure}
\end{equation}
Here 
\begin{align}
D_{xy}& =\frac{1}{N_{c}} \left \langle \text{Tr}\Big( \tilde{U}(\bm{x}_{\perp}) \tilde{U}^{\dagger}(\bm{y}_{\perp}) \Big) \right \rangle \, , \nonumber \\
Q_{xy;zw} & =\frac{1}{N_{c}} \left \langle \text{Tr} \Big( \tilde{U}(\bm{x}_{\perp}) \tilde{U}^{\dagger}(\bm{y}_{\perp})  \tilde{U}(\bm{z}_{\perp}) \tilde{U}^{\dagger}(\bm{w}_{\perp}) \Big)\right \rangle\, ,
\label{eq:dipole-quadrupole-Wilson-line-correlators}
\end{align} 
represent respectively the dipole and quadrupole Wilson line correlators, where 
\begin{equation}
\langle \hat{\mathcal{O}} \rangle = \int [ \mathcal{D} \rho_{A} ] \, W_{\Lambda_0^-} [\rho_{A}] \, \hat{\mathcal{O}} [\rho_{A}] \, ,
\label{eq:expectation-value-cgc}
 \end{equation}
denotes the expectation value of a generic operator $\hat{\mathcal{O}}$. The weight functional $W_{\Lambda_{0}^{-}} [\rho_{A}]$ contains fundamental information about $n$-body correlations amongst the color sources at the scale $\Lambda_{0}^{-}$. In the McLerran-Venugopalan model (MV)~\cite{McLerran:1993ni,McLerran:1993ka,McLerran:1994vd} where it was introduced, $W_{\Lambda_{0}^{-}} [\rho_{A}]$ is Gaussian distributed for a large nucleus~\cite{McLerran:1993ka,Kovchegov:1996ty,Jeon:2004rk} with a variance
$\mu^{2}_{A}\sim A^{1/3}$, where $A$ denotes the atomic number. In the MV model, 
$\mu_{A}^2\propto Q_{S,0}^2$, the saturation scale at $\Lambda_{0}^{-}$~\cite{Iancu:2003xm,Lappi:2007ku}. 
For quantitative estimates of the saturation scale at EIC energies, we refer the reader to \cite{Kowalski:2007rw,Aschenauer:2017jsk}. In the CGC EFT, $D$ and $Q$ appear in a variety of LO processes in both $p+A$ and $e+A$ collisions~\footnote{Group theory techniques to compute such correlators in the MV model are discussed in~\cite{Blaizot:2004wv,Dominguez:2012ad,Dusling:2017aot,Fukushima:2017mko}.}.

\begin{figure}[!htbp]
\begin{minipage}[b]{0.3\textwidth}
\includegraphics[width=\textwidth]{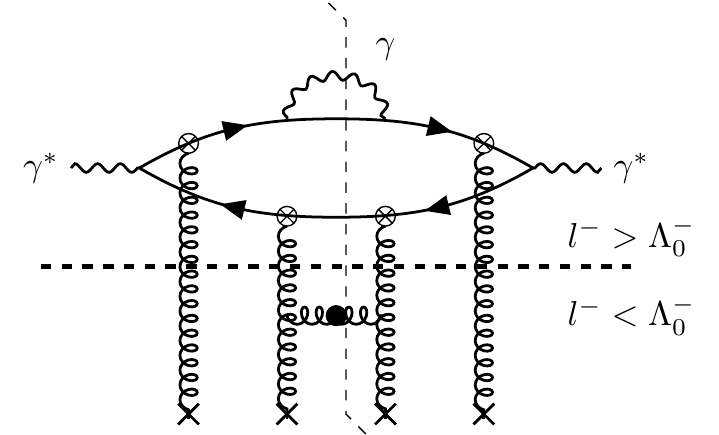}
\end{minipage}
\caption{NLO leading logs in $x$ ($\alpha_S \ln(1/x)\sim 1$) diagram with the same O(1) contributions as the LO diagram in Fig.~\ref{fig:one}. The filled blob represents the dressed gluon propagator in $A^-=0$ gauge. See text for details.\label{fig:two}}
\end{figure}

We turn now to the extension of our computation of $\tilde{X}_{\mu \nu}$ to NLO, details of which are provided in~\cite{Roy:2019hwr}. Let us first consider the NLO diagram in Fig.~\ref{fig:two}. 
An important ingredient in our computation is the gluon ``small fluctuations" propagator in the $A^-=0$ gauge classical shock wave background field~\cite{McLerran:1994vd,Ayala:1995hx,Ayala:1995kg,Balitsky:2001mr,Roy:2018jxq}: 
\begin{equation}
G_{\mu \nu;ab}(p,q) =  G^{0}_{\mu \rho;ac}(p)\,\mathcal{T}_{g}^{\rho \sigma;cd}(p,q)\,G^{0}_{\sigma \nu;db} (q) \, ,
\label{eq:dressed-gluon-mom-prop}
\end{equation}
where $G^{0}_{\mu \rho;ac}(p)=\frac{i}{p^{2}+i\varepsilon}\Big( -g_{\mu \rho}+\frac{p_{\mu}n_{\rho}+n_{\mu}p_{\rho}}{n.p} \Big)\delta_{ac}$ is the free propagator with  Lorentz indices 
$\mu,\rho$, color indices $a,c$ and $n^{\mu}=\delta^{\mu +}$ .
The effective gluon vertex 
\begin{align}
&\mathcal{T}^{\mu \nu;ab}_{g}(p,p')  =-2\pi \delta(p^{-}-p'^{-}) \times (2p^{-}) g^{\mu \nu} \, \text{sign}(p^{-}) \nonumber \\
& \times \int \mathrm{d}^{2} \bm{z}_{\perp} \, e^{-i(\bm{p}_{\perp}-\bm{p'}_{\perp}).\bm{z}_{\perp} }\,\, \Big( U^{ab} \Big)^{\text{sign}(p^{-})} (\bm{z}_{\perp}) \, ,
\label{eq:vertex-gluon}
\end{align}
corresponding to multiple scattering of the gluon off the shock wave background field, is represented by the filled blobs in Fig.~\ref{fig:two}. 

In the NLO diagrams represented in Fig.~\ref{fig:two}, the contributions enhanced by $\alpha_S \ln(\Lambda_1^-/\Lambda_0^-)$ (with 
$\Lambda_1^-$ chosen such that these terms are O(1)) can be combined with the LO result in Eq.~(\ref{eq:leading-order-dsigma}) and expressed as~\footnote{We employ here the Hermiticity of $W$ with respect to the functional integration over $\rho_{A}$.},
\begin{align}
\tilde{X}_{\mu \nu}^{\text{LO}} + \delta \tilde{X}_{\mu \nu}^{\text{NLO:1}} &=\!\! \int [\mathcal{D}\rho_{A}] \Big(  1+ \ln(\Lambda_{1}^{-}/ \Lambda_{0}^{-}) \mathcal{H}_{\text{LO}} \Big) W_{\Lambda^{-}_{0}} [\rho_{A}]  \nonumber \\
& \times  \hat{X}_{\mu \nu}^{\text{LO}} [\rho_{A}] \, .
\label{eq:LO-JIMWLK}
\end{align}
Further redefining
\begin{equation}
 \Big( 1+ \ln(\Lambda_{1}^{-}/ \Lambda_{0}^{-}) \mathcal{H}_{\text{LO}} \Big) W_{\Lambda_{0}^{-}}[\rho_{A}] = W_{\Lambda_{1}^{-}}[\rho_{A}] \, ,
 \label{eq:W-LLx}
\end{equation}
and thereby absorbing the semi-fast gluon fluctuations of the target in a modification of the weight functional of the color sources at the scale $\Lambda_1^-$, one obtains the leading log in $x$ (LL$x$)~\footnote{The  LC momentum fraction $x$ is equated to the Bjorken variable $x_{\rm Bj}$ here. The precise relation between the two is established beyond this order of the computation.} JIMWLK renormalization group equation~\footnote{To LL$x$, this equation generates the Balitsky hierarchy~\cite{Balitsky:1995ub} describing the evolution of n-point Wilson line correlators in $x$. In the limit of large number of colors $N_c$, and for $A \gg 1$, the simplest dipole correlator of light-like Wilson lines in this hierarchy satisfies the Balitsky-Kovchegov (BK) equation~\cite{Balitsky:1995ub,Kovchegov:1999yj}. In the leading twist limit where $Q_S^2 (x) /Q^2 \ll 1$, it reduces to the BFKL equation~\cite{Kuraev:1977fs,Balitsky:1978ic}.},
\begin{equation}
\frac{\partial}{\partial(\ln \Lambda^{-})} W_{\Lambda^{-}} [\rho_{A}] = \mathcal{H}_{\text{LO}}\, W_{\Lambda^{-}} [\rho_{A}] \, ,
\label{eq:LO-JIMWLK_final}
\end{equation}
where $\mathcal{H}_{\text{LO}}$ is the well-known JIMWLK Hamiltonian~\cite{JalilianMarian:1997gr,JalilianMarian:1997dw,Iancu:2000hn,Ferreiro:2001qy}. 
We will henceforth label the weight functional that satisfies Eq.~(\ref{eq:LO-JIMWLK_final}) as $W^{LLx}[\rho_{A}]$.

\begin{figure}[!htbp]
\begin{minipage}[b]{0.52\textwidth}
\includegraphics[width=\textwidth]{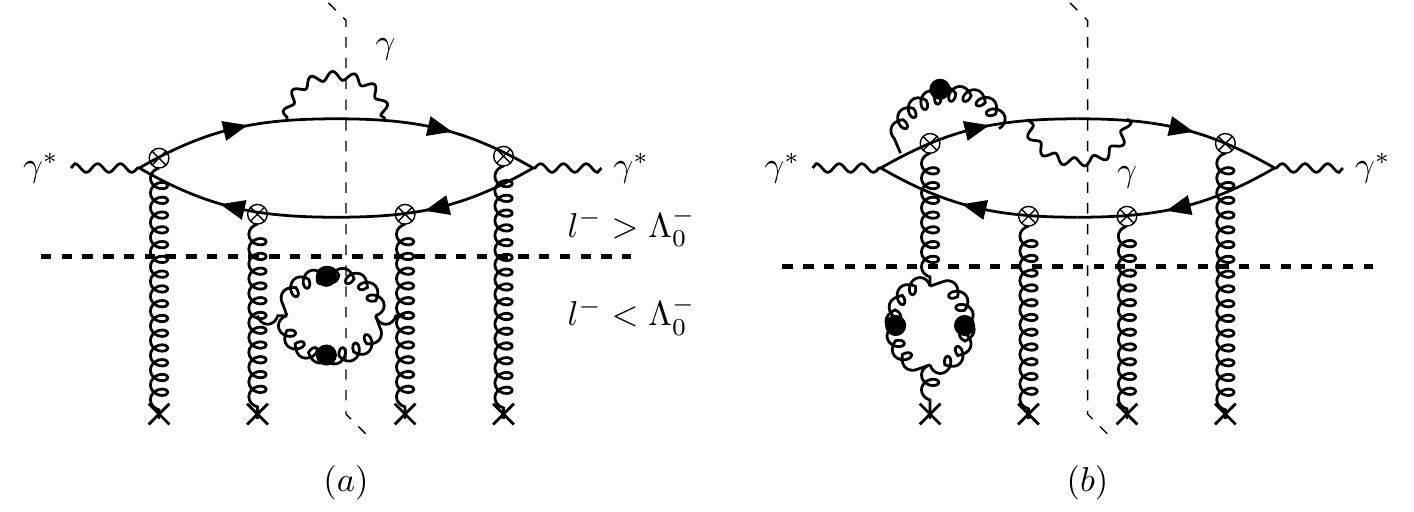}
\end{minipage}
\caption{(a) NNLO diagram  
corresponding to target fluctuations of O($\alpha_S^2 \ln(1/x)$). Such contributions are resummed to all orders by the NLO JIMWLK equation. (b) NNLO diagram with O($\alpha_S \ln(1/x)$) fluctuations of the target and O($\alpha_S$) fluctuations of the projectile. These projectile fluctuations constitute the NLO $\gamma$+dijet impact factor.\label{fig:three}}
\end{figure}

At next-to-next-to-leading order (NNLO) in $\alpha_S$, there are two relevant classes of contributions as illustrated in Fig.~\ref{fig:three}. Diagrams corresponding to a two loop fluctuation of the target are shown in Fig.~\ref{fig:three}(a). For such two loop diagrams, contributions~\footnote{There are also contributions from  two-loop QCD diagrams proportional to $\alpha_{S}^{2}$ alone (without leading logs in $x$) but these are suppressed at the desired accuracy of our problem.} of order $\alpha_{S}^{2} \ln^2(\Lambda_{1}^{-}/\Lambda_{0}^{-})\sim O(1)$ are included in $W^{LLx}[\rho_{A}]$. We will therefore consider here only the two loop diagrams that contain next-to-leading logarithms in $x$ (NLL$x$) contributions to Eq.~\ref{eq:LO-JIMWLK}. The LO+NLL$x$ result including these can be expressed as
\begin{equation}
\tilde{X}_{\mu \nu}^{\text{LO}} + \delta \tilde{X}_{\mu \nu}^{\text{NLL$x$}}   = \int      [\mathcal{D}\rho_{A}] \,  W_{\Lambda_1^-}^{NLLx}  [\rho_{A}]  \hat{X}_{\mu \nu}^{\text{LO}} [\rho_{A}] \, ,
\label{eq:NLL-JIMWLK}
\end{equation}
where
\begin{equation}
W_{\Lambda_1^-}^{NLLx}  [\rho_{A}] =\Big\{  1+ \ln(\Lambda_{1}^{-}/ \Lambda_{0}^{-}) (\mathcal{H}_{\text{LO}}+\mathcal{H}_{\text{NLO}})  \Big\}\,W_{\Lambda_{0}^{-}}[\rho_{A}]\, ,
\label{eq:NLL-WF}
\end{equation}
and the NLO JIMWLK Hamiltonian $\mathcal{H}_{\text{NLO}}$ computed in \cite{Balitsky:2013fea,Kovner:2013ona,Balitsky:2014mca,Lublinsky:2016meo,Caron-Huot:2013fea} (see also  \cite{Kovchegov:2006vj,Braun:2007vi}) is of order $\alpha_{S}^{2}$.

The second class of NNLO contributions (formally of O($\alpha_S^2$)), shown in Fig.~\ref{fig:three}(b), correspond to one loop fluctuations of both the projectile and the target. Specifically, the $\alpha_S\ln(1/x)$ contributions from the gluon fluctuations below the cut $\Lambda_{0}^{-}$ are matched to the finite contributions above the cut (without logarithms) of O($\alpha_S$) in the real and virtual corrections to the LO photon+dijet projectile final state. These finite terms constitute the NLO $\gamma$+dijet impact factor. Together, they give for the class of NNLO contributions in Fig.~\ref{fig:three}(b), 
\begin{equation}
\tilde{X}_{\mu \nu}^{\text{NNLO;finite}}  =  \int [\mathcal{D} \rho_{A}] \,  W^{LLx}[\rho_{A}] \, \hat{X}_{\mu \nu}^ {\text{NLO;finite}} [\rho_{A}]  \,.
\label{eq:LOJIMWLK-NLO}
\end{equation}



\begin{figure}[!htbp]
\begin{minipage}[b]{0.3\textwidth}
\includegraphics[width=\textwidth]{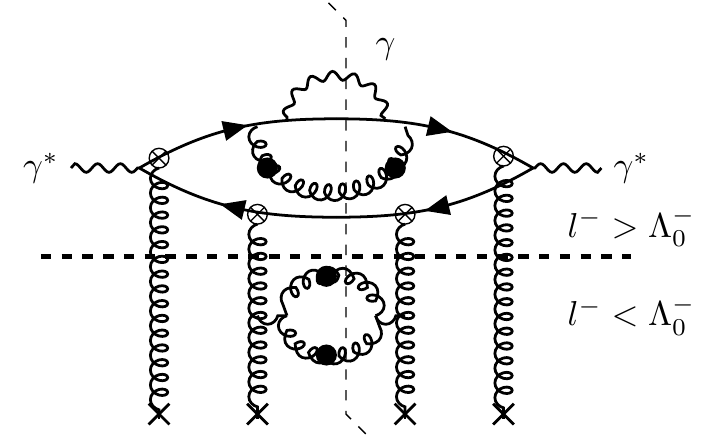}
\end{minipage}
\caption{N$^3$LO diagrams including O($\alpha_S$) fluctuations of the projectile and O($\alpha_S^2\ln(1/x)$) fluctuations of the target. Such contributions can be computed using extant results for NLO JIMWLK and the NLO impact factor computed in \cite{Roy:2019hwr}. \label{fig:five}}
\end{figure}

Combining the expressions in Eqs.~\ref{eq:NLL-JIMWLK} and \ref{eq:LOJIMWLK-NLO}, the hadron tensor for inclusive photon+dijet production to NLO+NLL$x$ accuracy can be written as
\begin{align}
& {\tilde X}_{\mu\nu}^{\text{NLO}+\text{NLL$x$}} = \!\!  \int [\mathcal{D} \rho_{A}] \, \Big\{ W^{NLLx}[\rho_{A}]\, \hat{X}_{\mu \nu}^{\text{LO}} [\rho_{A}]   \nonumber \\
&+ W^{LLx}[\rho_{A}] \, \hat{X}_{\mu \nu}^ {\text{NLO;finite}} [\rho_{A}] \Big\}  \nonumber \\
& \simeq  \int [\mathcal{D} \rho_{A}] \,  \Big( W^{NLLx} [\rho_{A}] \, \Big\{ \hat{X}_{\mu \nu}^{\text{LO}} [\rho_{A}] + \hat{X}_{\mu \nu}^ {\text{NLO;finite}} [\rho_{A}]  \Big\}  \nonumber \\
&+ O(\alpha_{S}^{3} \ln(\Lambda^{-}_{1}/\Lambda^{-}_{0}) ) \Big) \, .
\label{eq:dsigma-NLO-NLLx}
\end{align}
 Our knowledge of the NLO impact factor and NLL$x$ JIMWLK evolution can be combined, as shown above and in Fig.~\ref{fig:five}, to extend the scope of the computation to $O(\alpha_S^3 \ln(1/x))$. However as the $\simeq$ symbol indicates, this knowledge is insufficient to capture all the diagrams that contribute to this accuracy.   

 We shall now sketch the key features of the computation of the inclusive NLO photon+dijet impact factor in \cite{Roy:2019hwr}. 
The quantum fluctuations (with $l^{-}>\Lambda_{0}^{-}$)  that contribute towards the NLO impact factor can be broadly classified into the modulus squared of real gluon emission amplitudes and the interference of virtual gluon exchange processes with LO diagrams. In each case, the real or virtual gluon can scatter off the shock wave or propagate freely without scattering; in addition, there are all possible permutations of the emission vertex of the final state photon. For real emissions alone, there are 400 possible diagrams -- see~\cite{Roy:2019hwr} for the complete set of real and virtual graphs contributing at NLO. These can be categorized systematically by their color structures, allowing one to clearly observe the cancellation of the soft, collinear and ultraviolet (UV) divergences that arise in the intermediate steps of our computation. 

Soft singularities arise from the spurious $l^{-}=0$ pole in the free gluon propagator in $A^-=0$ LC gauge. These are regulated by imposing a cutoff at the initial scale of evolution $\Lambda^{-}_{0}$. We show in ~\cite{Roy:2019hwr} that log divergent terms in $\Lambda_{0}^{-}$ 
in the ``slow" gluon $l^{-} \rightarrow 0$ limit possess color structures at NLO  identical to those resulting from the action (as shown in~\cite{Dominguez:2011gc}) of the JIMWLK kernel on $\Xi$ in Eq.~\ref{eq:LO-cross-section-color-structure}. Our computation therefore provides an explicit proof of high energy JIMWLK factorization for a non-trivial process other than fully inclusive DIS. 

The UV divergences are extracted using dimensional regularization in $d=2-\epsilon$ dimensions. For gluon loops, most of the UV divergences cancel between graphs at the amplitude level. There 
are however residual collinear divergences. Such collinear singularities also arise from real gluon emission when integrating over the phase space in which the gluon can be collinear to the quark or antiquark. Because we are not integrating over the momenta of our $\gamma + q\bar{q}$ final state, there are collinear divergences that survive the real-virtual cancellations. These can be absorbed into a jet algorithm. Infrared (IR) safe quantities are obtained by promoting the partons to jets and working in the small cone approximation~\cite{Ivanov:2012ms} of jet cone radius $R \ll 1$. This restricts the integration over the phase space for the real gluon. The dominant contribution is of the form $\alpha_{S} (A \, \ln (R) +B)$, where $A, B$,  spelled out in \cite{Roy:2019hwr}, are of $O(1)$; all non-collinearly divergent contributions are phase space suppressed by powers of $R^{2}$. 

The jet algorithm also allows for a cancellation of soft-collinear divergences between soft gluon emissions inside and outside the jet cone. In the latter case, we observe that slow gluon emissions at wide angles ($l^{-} \rightarrow 0$ but any $l_{\perp}$), satisfy JIMWLK evolution and must be subtracted from the jet cross-section to avoid double counting when the NLO impact factor is combined with small $x$ evolution~\footnote{This correspondence is a feature of non-global logarithms in jet physics~\cite{Banfi:2002hw}; the latter was identified with BK/JIMWLK evolution by Marchesini and Mueller~\cite{Marchesini:2003nh} as well as by Weigert~\cite{Weigert:2003mm}, and subsequently significantly developed by Hatta et al.~\cite{Hatta:2008st,Avsar:2009yb,Hatta:2013iba,Hatta:2017fwr}. and by Neill~\cite{Neill:2016stq}.}. This result is an explicit realization of the conformal spacelike-timelike  correspondence noted previously by Mueller~\cite{Mueller:2018llt}.

We also observe interestingly that, as a consequence of the different topologies of the color structures that contribute towards soft and collinear divergences, the soft gluon theorem is violated for inclusive photon+dijet production~\footnote{This is unlike the case of diffractive DIS~\cite{Boussarie:2016ogo}.}. The factorization violating term has the color structure $(Q-DD)$. Since the building block of $Q$ and $D$ is the $x^-$ path ordered Wilson line in Eq.~\ref{eq:Wilson-line}, it would be interesting to 
explore if the soft gluon theorem can be restored by modifying the boundary conditions of the quadrupole and dipole operators at $x^-=\pm \infty$~\footnote{Since the soft gluon theorem is related to an infinite dimensional Kac-Moody symmetry~\cite{He:2015zea} on the celestial sphere (obtained by a stereographic projection of transverse coordinates) at null infinity, these symmetries may help identify the correct boundary conditions. Note that this  theorem is associated with a color memory~\cite{Pate:2017vwa}. In the Regge limit, the latter is precisely the color matrix in Eq.~\ref{eq:Wilson-line}~\cite{Ball:2018prg} at $x^-=\pm\infty$.}. 

After due consideration of all divergences, our final result for the triple differential cross-section for the $\gamma+q\bar{q}$ jet production in e+A DIS is 
\begin{align}
\frac{\mathrm{d}^{3} \sigma^{\text{LO+NLO+NLL$x$};\rm jet}}{\mathrm{d}x \mathrm{d}Q^{2} \mathrm{d}^{6} K_{\perp} \mathrm{d}^{3} \eta_{K} }& = \frac{\alpha_{em}^{2}q_{f}^{4}y^{2}N_{c}}{512 \pi^{5} Q^{2}} \, \frac{1}{(2\pi)^{4}} \,  \frac{1}{2} \nonumber \\
& \times L^{\mu \nu}  {\tilde X}_{\mu \nu}^{\text{NLO+NLL$x$}; \rm jet}  \,,
\end{align}
where the hadron tensor at $O(\alpha_{S}^{3} \ln (1/x))$ accuracy can be written as
\begin{widetext}
\begin{align}
{\tilde X}_{\mu \nu}^{\text{NLO+NLL$x$}; \rm jet} =   \int [\mathcal{D} \rho_A] \, W_{x_{\rm Bj}}^{ NLLx}[\rho_A] \Bigg[\Bigg( 1+  \frac{2\alpha_{S}C_{F}}{\pi} \, \Bigg\{ -\frac{3}{4} \ln \Big(  \frac{R^{2} \vert \bm{p}_{J\perp} \vert \, \vert \bm{p}_{K\perp} \vert  }{4 z_{J} z_{K} Q^{2} e^{\gamma_{E}}} \Big) +\frac{7}{4} -\frac{\pi^{2}}{6} \Bigg\} \, \Bigg) {\tilde X}_{\mu\nu}^{\rm LO; jet} [\rho_{A}]  + {\tilde X}_{\mu \nu; \rm finite}^{\rm NLO; \rm jet} [\rho_{A}] \, \Bigg] \,.
\label{eq:Xmunu-final}
\end{align}
\end{widetext}
In this expression~\footnote{$\bm{p}_{J \perp,K\perp}$ are the transverse momenta carried by the two jets and $z_{J},z_{K}$ are their respective momentum fractions relative to the projectile momentum. $\gamma_{E}$ is the Euler-Mascheroni constant.}, the finite terms ${\tilde X}_{\mu \nu; \rm finite}^{\rm NLO; jet} $ are of order $\alpha_S$ relative to the leading term and constitute the NLO impact factor. The explicit results for these are the principal results of~\cite{Roy:2019hwr}. For the virtual gluon diagrams, where the isolation of divergent and finite pieces is done at the amplitude level, it is straightforward albeit tedious to derive analytical expressions for such  terms. This is however not possible for real gluon emission graphs; we need to evaluate the finite pieces numerically. These are obtained by taking the modulus squared of the real emission amplitudes, integrating over the gluon phase space with a cutoff, implementing the jet algorithm, and subsequently subtracting the pieces that contribute to leading log JIMWLK evolution.

The numerical computation of the finite pieces constituting  ${\tilde X}_{\mu\nu; \rm finite}^{\rm NLO; jet}$, along with NLO BK/JIMWLK evolution, provide the necessary ingredients to compute photon+dijet production (and associated measurement channels) in e+A DIS to O($\alpha_S^3  \ln (1/x)$) accuracy. Prior NLO studies on DIS at small $x$ focused on the cross-section for fully inclusive DIS~\cite{Bartels:2000gt,Bartels:2002uz,Bartels:2001mv,Balitsky:2010ze,Balitsky:2012bs,Beuf:2011xd,Beuf:2016wdz,Beuf:2017bpd,Hanninen:2017ddy},  a noteworthy exception being the 
NLO studies of diffractive dijet and exclusive vector meson production by Boussarie et al.~\cite{Boussarie:2014lxa,Boussarie:2016ogo,Boussarie:2016bkq}. In this regard, our work goes a step beyond by considering more differential final states. The realization of these precision studies, while a formidable task, is feasible and will pave the way towards the quantitative global analyses of data required to uncover definitive evidence of gluon saturation. 


We note finally that the simple forms of the momentum space shockwave propagators in $A^- = 0$ gauge and the momentum space techniques employed in our work~\cite{Roy:2018jxq,Roy:2019hwr} will allow us to extend the accuracy of our computation to two loops. It is also worth mentioning that our NLO real gluon emission computation contains the LO results for the production cross-sections for 4-jet $\gamma+q\bar{q}g$ and 3-jet $q\bar{q}g$~\cite{Ayala:2016lhd} final states at small $x$.
 One may also consider employing this framework in p+A collisions, beyond the current state-of-the art for inclusive hadron~\cite{Chirilli:2012jd,Chirilli:2011km,Stasto:2011ru,Altinoluk:2014eka,Ducloue:2019ezk}, quarkonium~\cite{Kang:2013hta,Ma:2014mri} and photon production~\cite{Benic:2016uku,Benic:2016yqt,Benic:2018hvb}, to NNLO in the CGC power counting and beyond. 

This material is based on work supported by the U.S. Department of Energy, Office of Science, Office of Nuclear Physics, under Contracts No. de-sc0012704 and within the framework of the TMD Theory Topical Collaboration. K. R is supported by an LDRD grant from Brookhaven Science Associates and by the Joint BNL-Stony Brook Center for Frontiers in Nuclear Science (CFNS).

\bibliography{bibliography.bib}

\end{document}